\documentclass[12pt]{article}
\pdfoutput=1
\usepackage{latexsym, graphicx,cite} 
\usepackage{amsmath}
\usepackage{amssymb}

 \newcommand{\arXiv}[1]{\href{http://www.arXiv.org/abs/#1}{#1}}
\usepackage[colorlinks=true, linkcolor=blue, bookmarks=true]{hyperref}

\makeatletter
\renewcommand\section{\@startsection {section}{1}{\z@}%
                                   {-3.5ex \@plus -1ex \@minus -.2ex}
                                   {2.3ex \@plus.2ex}%
                                   {\normalfont\large\bfseries}}
\renewcommand\subsection{\@startsection{subsection}{2}{\z@}%
                                     {-3.25ex\@plus -1ex \@minus -.2ex}%
                                     {1.5ex \@plus .2ex}%
                                     {\normalfont\bfseries}}
\makeatother

\newcommand{\beq}{\begin{equation}}
\newcommand{\eeq}{\end{equation}}
\newcommand{\ber}{\begin{array}}
\newcommand{\eer}{\end{array}}

\newcommand{\eps}{\varepsilon}

\newcommand{\om}{\omega}

\newcommand{\ena}{\end{eqnarray}}
\newcommand{\beqa}{\begin{eqnarray}}
\newcommand{\eeqa}{\end{eqnarray}}
\newcommand{\bea}{\begin{eqnarray}}
\newcommand{\eea}{\end{eqnarray}}

\newcommand{\be}{\begin{equation}}
\newcommand{\ee}{\end{equation}}

\begin{document}
\begin{titlepage}
\begin{flushright}
\phantom{arXiv:yymm.nnnn}
\end{flushright}
\vfill
\begin{center}
{\LARGE\bf Ultraviolet asymptotics and singular\vspace{2mm}\\ dynamics of AdS perturbations}    \\
\vskip 15mm
{\large Ben Craps$^a$, Oleg Evnin$^{b,a}$ and Joris Vanhoof$^a$}
\vskip 7mm
{\em $^a$ Theoretische Natuurkunde, Vrije Universiteit Brussel and\\
The International Solvay Institutes\\ Pleinlaan 2, B-1050 Brussels, Belgium}
\vskip 3mm
{\em $^b$ Department of Physics, Faculty of Science, Chulalongkorn University,\\
Thanon Phayathai, Pathumwan, Bangkok 10330, Thailand}

\vskip 3mm
{\small\noindent  {\tt Ben.Craps@vub.ac.be, oleg.evnin@gmail.com, Joris.Vanhoof@vub.ac.be}}
\vskip 10mm
\end{center}
\vfill

\begin{center}
{\bf ABSTRACT}\vspace{3mm}
\end{center}

Important insights into the dynamics of spherically symmetric AdS-scalar field perturbations can be obtained by considering a simplified time-averaged theory accurately describing perturbations of amplitude $\eps$ on time-scales of order $1/\eps^2$. The coefficients of the time-averaged equations are complicated expressions in terms  of the AdS scalar field mode functions, which are in turn related to the Jacobi polynomials. We analyze the behavior of these coefficients for high frequency modes. The resulting asymptotics can be useful for understanding the properties of the finite-time singularity in solutions of the time-averaged theory recently reported in the literature. We highlight, in particular, the gauge dependence of this asymptotics, with respect to the two most commonly used gauges. The harsher growth of the coefficients at large frequencies in higher-dimensional AdS suggests strengthening of turbulent instabilities in higher dimensions. In the course of our derivations, we arrive at recursive relations for the coefficients of the time-averaged theory that are likely to be useful for evaluating them more efficiently in numerical simulations.

\vfill

\end{titlepage}


\section{Introduction}

Anti-de Sitter (AdS) space-time is linearly stable. Yet, non-linear dynamics of its perturbations is characterized by complex interplay of stable and unstable behavior. Numerical simulations for a massless, spherically symmetric scalar field \cite{BR,MRlec,DHMS,MRperiod,BLL,Abajo-Arrastia:2014fma} indicate that, while some initial data of small amplitude $\eps$ collapse, after multiple reflections from the AdS boundary, to form black holes on time scales of order $1/\eps^2$, other small amplitude initial data remain small for much longer (possibly, for all) times; see also \cite{Kim:2014ida, Okawa:2015xma, Deppe:2015qsa} for generalizations to massive scalar fields.

Attempts to control the small amplitude dynamics in terms of na\"\i ve perturbative expansions in the perturbation amplitude $\eps$ are plagued by a profusion of secular terms \cite{BR} growing in time and invalidating the perturbative expansions. This profusion is due to the perfectly commensurate spectrum of frequencies displayed by the linearized AdS perturbations. The perturbation theory can be re-structured using a variety of techniques equivalent at the lowest order (multiscale or renormalization group resummation, averaging), producing an improved perturbative expansion accurately describing the small amplitude dynamics on time scales of order $1/\eps^2$ \cite{FPU,CEV1,CEV2, BMR, GMLL}.

As a result of applying the resummation techniques in the small amplitude regime, one obtains a system of effective equations governing slow variations of the amplitudes and phases of the AdS normal modes under the effect of non-linearities. We usually refer to these equations as ``(renormalization) flow equations'' or ``time-averaged equations''; they have also been called ``two-time framework equations'' \cite{FPU} and ``resonant system''  \cite{BMR}. The coefficients in this system are complicated expressions in terms of integrals of the AdS mode functions and their derivatives, obtained in \cite{CEV1,CEV2}.

In \cite{BMR}, the flow equations were solved numerically for AdS$_5$. It was observed that, for a variety of initial data for which collapse occurs in the fully non-linear description, a singularity in solutions develops in finite time. At this singular moment, the time derivatives of the phases diverge logarithmically, while the amplitude spectrum, initially rapidly decreasing in the ultraviolet, attains a power-law form $1/n^2$, where $n$ is the mode number. This signifies efficient weakly turbulent transfer of energy to short wavelength modes, which is the main mechanism behind the collapse dynamics.

Since the key questions on the fate of AdS perturbations involve ultraviolet dynamics, it is important to develop asymptotic techniques for analyzing the high-frequency contributions to the flow equations. Here, we initiate such analysis by addressing the high-frequency asymptotics of the interaction coefficients appearing in the flow equations (see also our parallel work \cite{CEJV}).\footnote{Some further considerations of ultraviolet properties of AdS perturbations were given in \cite{MS} after the present article had appeared as a preprint.} To this end, we first review, in section 2, the basics of constructing the flow equations for spherically symmetric AdS-scalar field perturbations. In section 3, we develop iterative relations for the interaction coefficients in these equations. These iterative relations may be of independent value for the purposes of optimizing numerical simulations of the flow equations, as in \cite{BMR}. In section 4, we use the iterative relations to reveal the power law behavior of the interation coefficients for large mode numbers, and emphasize that this asymptotics is substantially affected by the time parametrization conventions (two gauges, corresponding to labelling the time slices by the proper time in the center and at the boundary of the perturbed AdS, have commonly appeared in the literature). We conclude with a discussion of the qualitative implications of the power law behavior we find, in particular, the qualitative dependence of the near-AdS dynamics on the number of spatial dimensions.

\section{Weakly non-linear AdS dynamics and the effective time-averaged description}

Reviews of the general spherically symmetric AdS-scalar field set-up are widely available in the literature, for example \cite{MRlec} or our own work \cite{CEV1,CEV2}. We shall therefore limit ourselves to a very basic summary of the relevant formulae.

One considers the following spherically symmetric metric:
\begin{equation}\label{eqn:MetricAnsatz}
ds^{2}=\frac{L^{2}}{\cos^{2}x}\left(\frac{dx^{2}}{A}-Ae^{-2\delta}dt^{2}+\sin^{2}x\,d\Omega_{d-1}^{2}\right),
\end{equation}
with  $A(x,t)$, $\delta(x,t)$ and a scalar field $\phi(x,t)$ depending on the time coordinate $t$ and the radial coordinate $x \in [0,\pi/2)$.   $A=1$, $\delta=\phi=0$ corresponds to empty AdS$_{d+1}$.
$\delta(x,t)$ has some residual gauge freedom $\delta(x,t)\mapsto\delta(x,t)+q(t)$, corresponding to redefinition of time. Two possible gauge fixing conditions have appeared in the literature: $\delta(0,t)=0$ \cite{BR, FPU, CEV1,BMR,GMLL} and $\delta(\pi/2,t)=0$ \cite{BLL, CEV2}. We shall refer to them as the `interior' and `boundary' time, respectively.

The equations of motion take the form
\begin{subequations}
\label{eqn:EOM}
\begin{align}
\dot{\Phi}&=\left(Ae^{-\delta}\Pi\right)',
&\dot{\Pi}&=\frac{1}{\mu}\left(\mu Ae^{-\delta}\Phi\right)', \\
A'&=\frac{\nu'}{\nu}\left(A-1\right)-\mu\nu\left(\Phi^{2}+\Pi^{2}\right)A,
&\delta'&=-\mu\nu\left(\Phi^{2}+\Pi^{2}\right), \label{eqn:EOMConstraint}
\end{align}
\begin{equation}
\dot{A}=-2\mu\nu A^{2}e^{-\delta}\Phi\Pi,
\end{equation}
\end{subequations}
with $\Phi\equiv\phi'$ and $\Pi\equiv A^{-1}e^{\delta}\dot{\phi}$ (where dots and primes denote the $t$- and $x$-derivatives, respectively),
\begin{equation}
\mu(x)\equiv(\tan x)^{d-1}
\qquad\text{and}\qquad
\nu(x)\equiv\frac{(d-1)}{\mu'(x)}=\frac{\sin x\cos x}{(\tan x)^{d-1}}.
\label{munu}
\end{equation}
These equations, as they are, are forbiddingly complicated. Further analytic progress can be made by considering small amplitude solutions. Na\"\i ve power law expansions in the field amplitude result, however, in unacceptable perturbation series plagued by growing `secular' terms \cite{BR,FPU,CEV1,CEV2}. Improved perturbative expansions can be constructed using multiscale \cite{FPU}, renormalization \cite{CEV1} and averaging \cite{CEV2} methods, all of which are equivalent at the lowest non-trivial order. The general idea is to use as the starting point for weak field expansions a linearized solution with slowly modulated integration constants, namely
\beq
\phi(x,t)=\eps\sum\limits_{n=0}^\infty A_n(\eps^2t)\, e_n(x) \cos(\om_n t+B_n(\eps^2 t)).
\eeq
Here, $e_n$ are the mode functions of a linearized massless scalar field in fixed AdS$_{d+1}$,
\beq
e_{n}(x)=k_{n}(\cos x)^{d}P_{n}^{\left(\frac{d}{2}-1,\frac{d}{2}\right)}\left(\cos(2x)\right)
\qquad\text{with}\qquad
k_{n}=\frac{2\sqrt{n!(n+d-1)!}}{\Gamma\left(n+\frac{d}{2}\right)}.
\label{modefn}
\end{equation}
$P_{n}^{(a,b)}(x)$ is a Jacobi polynomial of order $n$. $\om_n=d+2n$ is the famously integer, fully commensurate spectum of the linearized normal mode frequencies in AdS, largely responsible for the rich non-linear dynamics of AdS perturbations.

The effective equations for the slow time dependences of $A_n(\eps^2t)$ and $B_n(\eps^2t)$ can be deduced using a number of equivalent methods \cite{FPU,CEV1,CEV2}. Their analytic form was presented in \cite{CEV1,CEV2}. The equations are known to describe the exact non-linear dynamics accurately on time scales of order $1/\eps^2$, if the scalar field amplitude is of order $\eps$ \cite{CEV2}. Some phenomenological consequences of these equations have been explored in \cite{BMR,GMLL}.

The averaged equations \cite{CEV2} are of the form
\begin{align}\label{Eqn:RG}
\frac{2\omega_{l}}{\epsilon^{2}}\frac{dA_{l}}{dt}=&-\underbrace{\sum_{i}^{\{i,j\}}\sum_{j}^{\neq}\sum_{k}^{\{k,l\}}}_{\omega_{i}+\omega_{j}=\omega_{k}+\omega_{l}}S_{ijkl}A_{i}A_{j}A_{k}\sin(B_{l}+B_{k}-B_{i}-B_{j}), \\
\frac{2\omega_{l}A_{l}}{\epsilon^{2}}\frac{dB_{l}}{dt}=&-T_{l}A_{l}^{3}-\sum_{i}^{i\neq l}R_{il}A_{i}^{2}A_{l}-\underbrace{\sum_{i}^{\{i,j\}}\sum_{j}^{\neq}\sum_{k}^{\{k,l\}}}_{\omega_{i}+\omega_{j}=\omega_{k}+\omega_{l}}S_{ijkl}A_{i}A_{j}A_{k}\cos(B_{l}+B_{k}-B_{i}-B_{j}).
\end{align}
Note that {\em a priori} resonant frequency quartets other than $\om_i+\om_j=\om_k+\om_l$ might have appeared in the sum, but they do not in fact contribute \cite{CEV1} due to special selection rules in AdS (see also \cite{Yang,EK}). The coefficients of the terms that appear in the time-averaged equations can be expressed in terms of certain integrals of products of the mode functions. 

In boundary time gauge ($\delta(\pi/2)=0$), in which the slow dynamics turns out to be manifestly Hamiltonian \cite{CEV2}, the coefficients take the following form:
\begin{equation}
T_{l}=\frac{1}{2}\omega_{l}^{2}X_{llll}+\frac{3}{2}Y_{llll}+2\omega_{l}^{4}W^{(0,0)}_{llll}+2\omega_{l}^{2}W^{(1,0)}_{llll},
\end{equation}
\begin{align}
R_{il}=&
\frac{1}{2}\left(\frac{\omega_{i}^{2}+\omega_{l}^{2}}{\omega_{l}^{2}-\omega_{i}^{2}}\right)\left(\omega_{l}^{2}X_{illi}-\omega_{i}^{2}X_{liil}\right)+2\left(\frac{\omega_{l}^{2}Y_{ilil}-\omega_{i}^{2}Y_{lili}}{\omega_{l}^{2}-\omega_{i}^{2}}\right)+\frac{1}{2}(Y_{iill}+Y_{llii}) \nonumber \\
+&\left(\frac{\omega_{i}^{2}\omega_{l}^{2}}{\omega_{l}^{2}-\omega_{i}^{2}}\right)\left(X_{illi}-X_{lili}\right)+\omega_{i}^{2}\omega_{l}^{2}(W^{(0,0)}_{llii}+W^{(0,0)}_{iill})+\omega_{i}^{2}W^{(1,0)}_{llii}+\omega_{l}^{2}W^{(1,0)}_{iill},
\end{align}
\begin{align}
S_{ijkl}=&
-\frac{1}{4}\left(\frac{1}{\omega_{i}+\omega_{j}}+\frac{1}{\omega_{i}-\omega_{k}}+\frac{1}{\omega_{j}-\omega_{k}}\right)(\omega_{i}\omega_{j}\omega_{k}X_{lijk}-\omega_{l}Y_{iljk}) \nonumber \\
&-\frac{1}{4}\left(\frac{1}{\omega_{i}+\omega_{j}}+\frac{1}{\omega_{i}-\omega_{k}}-\frac{1}{\omega_{j}-\omega_{k}}\right)(\omega_{j}\omega_{k}\omega_{l}X_{ijkl}-\omega_{i}Y_{jikl}) \nonumber \\
&-\frac{1}{4}\left(\frac{1}{\omega_{i}+\omega_{j}}-\frac{1}{\omega_{i}-\omega_{k}}+\frac{1}{\omega_{j}-\omega_{k}}\right)(\omega_{i}\omega_{k}\omega_{l}X_{jikl}-\omega_{j}Y_{ijkl}) \nonumber \\
&-\frac{1}{4}\left(\frac{1}{\omega_{i}+\omega_{j}}-\frac{1}{\omega_{i}-\omega_{k}}-\frac{1}{\omega_{j}-\omega_{k}}\right)(\omega_{i}\omega_{j}\omega_{l}X_{kijl}-\omega_{k}Y_{ikjl}). \label{Eqn:Sijkl}
\end{align}
The integrals that appear in these expressions are defined by
\begin{subequations}
\begin{align}
X_{ijkl}&=\int_{0}^{\pi/2}\text{d}x\,e'_{i}(x)e_{j}(x)e_{k}(x)e_{l}(x)(\mu(x))^{2}\nu(x), \\
Y_{ijkl}&=\int_{0}^{\pi/2}\text{d}x\,e'_{i}(x)e_{j}(x)e'_{k}(x)e'_{l}(x)(\mu(x))^{2}\nu(x), \\
W^{(0,0)}_{ijkl}&=\int_{0}^{\pi/2}\text{d}x\,e_{i}(x)e_{j}(x)\mu(x)\nu(x)\int_{0}^{x}\text{d}y\,e_{k}(y)e_{l}(y)\mu(y), \label{W00}\\
W^{(1,0)}_{ijkl}&=\int_{0}^{\pi/2}\text{d}x\,e'_{i}(x)e'_{j}(x)\mu(x)\nu(x)\int_{0}^{x}\text{d}y\,e_{k}(y)e_{l}(y)\mu(y).\label{W10}
\end{align}
\end{subequations}
Note that the coefficient $R_{ij}=R_{ji}$ is symmetric. Whenever the resonance condition $\omega_{i}+\omega_{j}=\omega_{k}+\omega_{l}$ is satisfied, one also has $S_{ijkl}=S_{jikl}$ and $S_{ijkl}=S_{klij}$.

In the interior time gauge ($\delta(0,t)=0$), the expressions for $T_{i}$ and $R_{ij}$ are slightly more complicated  \cite{CEV1,CEV2}, with the additional terms $\omega^{2}_{i}(A_{ii}+\omega^{2}_{i}V_{ii})$ and $\omega^{2}_{j}(A_{ii}+\omega^{2}_{i}V_{ii})$, respectively. The $V$ and $A$ coefficients are defined as.
\begin{align}
V_{ij}&=\int_{0}^{\pi/2}\text{d}x\,e_{i}(x)e_{j}(x)\mu(x)\nu(x), &
A_{ij}&=\int_{0}^{\pi/2}\text{d}x\,e'_{i}(x)e'_{j}(x)\mu(x)\nu(x).
\label{VAdef}
\end{align}
Note that the difference between the two gauges only affects the equation for the phases $B_n$, but not the equations for the amplitudes $A_n$. In this gauge, the $R$ coefficients are not symmetric and the slow dynamics is not manifestly Hamiltonian \cite{CEV2}.

\section{Recursive analysis of the interaction coefficients}

\subsection{Recurrence relations for the mode functions}

We first analyze the mode-number structure of the AdS$_{d+1}$ mode functions given by (\ref{modefn}).
Using the recurrence relations for the Jacobi polynomials $P_{n}^{(\alpha,\beta)}(x)$,
\begin{align}
&2(n+1)(n+\alpha+\beta+1)(2n+\alpha+\beta)P_{n+1}^{(\alpha,\beta)}(x)=-2(n+\alpha)(n+\beta)(2n+\alpha+\beta+2)P_{n-1}^{(\alpha,\beta)}(x) \nonumber \\
&\qquad\qquad\quad+(2n+\alpha+\beta+1)\left\{(2n+\alpha+\beta+2)(2n+\alpha+\beta)x+\alpha^{2}-\beta^{2}\right\}P_{n}^{(\alpha,\beta)}(x)
\end{align}
and the derivative formula,
\begin{align}
&(2n+\alpha+\beta+2)(1-x^{2})\frac{d}{dx}P^{(\alpha,\beta)}_{n}(x)=-2(n+1)(n+\alpha+\beta+1)P^{(\alpha,\beta)}_{n+1}(x) \nonumber \\
&\qquad\qquad\qquad\qquad+(n+\alpha+\beta+1)\left(\alpha-\beta+(2n+\alpha+\beta+2)x\right)P^{(\alpha,\beta)}_{n}(x)
\end{align}
we can show that the mode functions satisfy the following identities,
\begin{align}
\mu(x)\nu'(x)e_{n}(x)=-(d-1)\left(1-\frac{1}{\omega_{n}^{2}-1}\right)e_{n}(x)&+\sqrt{(n+1)(n+d)}\left(\frac{1}{\omega_{n}+1}\right)e_{n+1}(x) \nonumber \\
&+\sqrt{n(n+d-1)}\left(\frac{1}{\omega_{n}-1}\right)e_{n-1}(x)
\label{Eqn:Recurrence1e}
\end{align}
and
\begin{align}
2\mu(x)\nu(x)E_{n}(x)=-(d-1)\left(\frac{\omega_{n}}{\omega_{n}^{2}-1}\right)e_{n}(x)&+\sqrt{(n+1)(n+d)}\left(\frac{1}{\omega_{n}+1}\right)e_{n+1}(x) \nonumber \\
&-\sqrt{n(n+d-1)}\left(\frac{1}{\omega_{n}-1}\right)e_{n-1}(x).
\label{Eqn:Recurrence2e}
\end{align}
We have introduced the notation $E_{n}(x)=(1/\omega_{n})e'_{n}(x)$. One can show that $E_{n}(x)$ is also an orthonormal family of functions with weight $\mu(x)$. By taking derivatives of these and using $-(\mu e'_{n})'=\omega_{n}^{2}\mu e_{n}$ and $-(\mu\nu')'=4\mu\nu$, we can obtain the relations
\begin{align}
\mu(x)\nu'(x)E_{n}(x)=&-(d-1)\left(1+\frac{1}{\omega_{n}^{2}-1}\right)E_{n}(x)+\sqrt{(n+1)(n+d)}\left(\frac{1}{\omega_{n}+1}\right)E_{n+1}(x) \nonumber \\
&+\sqrt{n(n+d-1)}\left(\frac{1}{\omega_{n}-1}\right)E_{n-1}(x)
\label{Eqn:Recurrence1E}
\end{align}
and
\begin{align}
2\mu(x)\nu(x)e_{n}(x)=&-(d-1)\left(\frac{\omega_{n}}{\omega_{n}^{2}-1}\right)E_{n}(x)-\sqrt{(n+1)(n+d)}\left(\frac{1}{\omega_{n}+1}\right)E_{n+1}(x) \nonumber \\
&+\sqrt{n(n+d-1)}\left(\frac{1}{\omega_{n}-1}\right)E_{n-1}(x).
\label{Eqn:Recurrence2E}
\end{align}
Note that up to a few signs, these recurrence relations preserve their form under interchange of $e_{n}(x)$ and $E_{n}(x)$. This feature will allow us to streamline some of our derivations below.

\subsection{Recurrence relations for the $X$ coefficients}

A useful simplification is attained by expressing the coefficients
\begin{equation}
X_{ijkl}=\int_{0}^{\frac{\pi}{2}}\text{d}x(\mu(x)\nu(x)e'_{i}(x))e_{j}(x)e_{k}(x)e_{l}(x)\mu(x)
\end{equation}
in terms of the totally symmetric combinations
\begin{equation}
\chi_{ijkl}=\int_{0}^{\frac{\pi}{2}}\text{d}x\,e_{i}(x)e_{j}(x)e_{k}(x)e_{l}(x)\mu(x)
\label{defchi}
\end{equation}
with the help of the recurrence relation \eqref{Eqn:Recurrence2e}. This results in the formula
\begin{align}
\frac{1}{\omega_{n}}X_{nmpq}=-\frac{1}{2}(d-1)\left(\frac{\omega_{n}}{\omega_{n}^{2}-1}\right)\chi_{nmpq}&+\frac{1}{2}\sqrt{(n+1)(n+d)}\left(\frac{1}{\omega_{n}+1}\right)\chi_{(n+1)mpq} \nonumber \\
&-\frac{1}{2}\sqrt{n(n+d-1)}\left(\frac{1}{\omega_{n}-1}\right)\chi_{(n-1)mpq}.
\label{Eqn:XfromChi}
\end{align}

We shall now derive a set of recursive relations for $\chi$. Once $\chi$ have been determined, $X$ can be extracted using the above formula.

Consider first the product $\mu\nu'e_ne_m$. One can apply the recurrence relation (\ref{Eqn:Recurrence1e}) either to $\mu\nu'e_n$ or to $\mu\nu'e_m$, resulting in two different re-writings of this product. Equating these two different re-writings, multiplying by $\mu e_pe_q$ and integrating over $x$, one obtains
\begin{align}
&(d-1)\left(\frac{\omega_{m}^{2}}{\omega_{m}^{2}-1}\right)\chi_{nmpq}+\sqrt{(m+1)(m+d)}\left(\frac{\chi_{n(m+1)pq}}{\omega_{m}+1}\right)+\sqrt{m(m+d-1)}\left(\frac{\chi_{n(m-1)pq}}{\omega_{m}-1}\right) \nonumber \\
&=(d-1)\left(\frac{\omega_{n}^{2}}{\omega_{n}^{2}-1}\right)\chi_{nmpq}+\sqrt{(n+1)(n+d)}\left(\frac{\chi_{(n+1)mpq}}{\omega_{n}+1}\right)+\sqrt{n(n+d-1)}\left(\frac{\chi_{(n-1)mpq}}{\omega_{n}-1}\right).
\label{Eqn:WaveEquationChi}
\end{align}
(By total symmetry of $\chi$, a similar relation holds for any other pair of indices.)

Furthermore, one can consider the relation
\beq\label{Eqn:IntegralChiN}
\int_{0}^{\pi/2}\text{d}x\,\mu^{2}\nu'e_{n}e_{m}e_{p}e_{q}=-2(d-1)\chi_{nmpq}-X_{nmpq}-X_{mnpq}-X_{pnmq}-X_{qnmp}.
\eeq
Here we have used integration by parts moving the derivative off $\nu'$, and the fact that $\mu'\nu=(d-1)$. This relation can be easily converted in a closed iterative relation for $\chi$ alone. On the right-hand side, we can eliminate $X$ in favor of $\chi$ with \eqref{Eqn:XfromChi}. On the left-hand side, applying (\ref{Eqn:Recurrence1e}) to $\mu\nu'e_n$ re-expresses everything through $\chi$ alone. The result is
\begin{align}
&\left\{\frac{1}{2}(d-1)\left(\frac{\omega_{n}^{2}}{\omega_{n}^{2}-1}\right)\chi_{nmpq}-\frac{1}{2}\sqrt{(n+1)(n+d)}\left(\frac{\omega_{n}}{\omega_{n}+1}\right)\chi_{(n+1)mpq}\right. \nonumber \\
&\left.+\frac{1}{2}\sqrt{n(n+d-1)}\left(\frac{\omega_{n}}{\omega_{n}-1}\right)\chi_{(n-1)mpq}\right\}+\{n\leftrightarrow m\}+\{n\leftrightarrow p\}+\{n\leftrightarrow q\} \nonumber \\
&=(d-1)\left(\frac{\omega_{n}^{2}}{\omega_{n}^{2}-1}\right)\chi_{nmpq}+\sqrt{(n+1)(n+d)}\left(\frac{1}{\omega_{n}+1}\right)\chi_{(n+1)mpq} \nonumber \\
&+\sqrt{n(n+d-1)}\left(\frac{1}{\omega_{n}-1}\right)\chi_{(n-1)mpq}.
\label{Eqn:HomogeneousEquationChi}
\end{align}
We shall now present the analogous relations for the $Y$-coefficients, and then explain how (\ref{Eqn:WaveEquationChi}) and (\ref{Eqn:HomogeneousEquationChi}), and the corresponding equations pertinent to the $Y$-coefficients, can be used to set up effective recursive evaluation algorithms.

\subsection{Recurrence relations for the $Y$ coefficients}

We have already remarked on the immediate parallels between the recurrence relations for $e_{n}(x)$ and $E_{n}(x)=(1/\omega_{n})e'_{n}(x)$. These parallels make the analysis of the $Y$ coefficients essentially identical to what we have presented above for the $X$ coefficients. Using recursion formula \eqref{Eqn:Recurrence2E}, we can write
\begin{equation}
Y_{ijkl}=\int_{0}^{\frac{\pi}{2}}\text{d}x\,e'_{i}(x)e_{j}(x)e'_{k}(x)e'_{l}(x)(\mu(x))^{2}\nu(x)
\end{equation}
in terms of
\begin{equation}
\psi_{ijkl}=\int_{0}^{\frac{\pi}{2}}\text{d}x\,E_{i}(x)E_{j}(x)E_{k}(x)E_{l}(x)\mu(x)
\end{equation}
as follows:
\begin{align}
\frac{1}{\omega_{n}\omega_{p}\omega_{q}}Y_{nmpq}=-\frac{1}{2}(d-1)\left(\frac{\omega_{m}}{\omega_{m}^{2}-1}\right)\psi_{nmpq}&-\frac{1}{2}\sqrt{(m+1)(m+d)}\left(\frac{1}{\omega_{m}+1}\right)\psi_{n(m+1)pq} \nonumber \\
&+\frac{1}{2}\sqrt{m(m+d-1)}\left(\frac{1}{\omega_{m}-1}\right)\psi_{n(m-1)pq}
\label{Eqn:YfromPsi}
\end{align}
Considering the action of \eqref{Eqn:Recurrence1E} on $\mu\nu'E_{n}E_{m}$ results in
\begin{align}
&-(d-1)\left(\frac{1}{\omega_{n}^{2}-1}\right)\psi_{nmpq}+\sqrt{(n+1)(n+d)}\left(\frac{\psi_{(n+1)mpq}}{\omega_{n}+1}\right)+\sqrt{n(n+d-1)}\left(\frac{\psi_{(n-1)mpq}}{\omega_{n}-1}\right) \nonumber \\
&=-(d-1)\left(\frac{1}{\omega_{m}^{2}-1}\right)\psi_{nmpq}+\sqrt{(m+1)(m+d)}\left(\frac{\psi_{n(m+1)pq}}{\omega_{m}+1}\right)+\sqrt{m(m+d-1)}\left(\frac{\psi_{n(m-1)pq}}{\omega_{m}-1}\right)
\label{Eqn:WaveEquationPsi}
\end{align}
On the other hand, one obtains by integration by parts
\beq
\int_{0}^{\pi/2}\text{d}x\,\mu^{2}\nu'E_{n}E_{m}E_{p}E_{q}=2(d-1)\psi_{nmpq}+\frac{\omega_{n}Y_{mnpq}}{\omega_{m}\omega_{p}\omega_{q}}+\frac{\omega_{m}Y_{nmpq}}{\omega_{n}\omega_{p}\omega_{q}}+\frac{\omega_{p}Y_{npmq}}{\omega_{n}\omega_{m}\omega_{q}}+\frac{\omega_{q}Y_{nqmp}}{\omega_{n}\omega_{m}\omega_{p}}
\eeq
Using  \eqref{Eqn:Recurrence1E} and \eqref{Eqn:YfromPsi}, we can conclude that
\begin{align}
&\left\{-\frac{1}{2}(d-1)\left(\frac{\omega_{n}^{2}}{\omega_{n}^{2}-1}\right)\psi_{nmpq}-\frac{1}{2}\sqrt{(n+1)(n+d)}\left(\frac{\omega_{n}}{\omega_{n}+1}\right)\psi_{(n+1)mpq}\right. \nonumber \\
&\left.+\frac{1}{2}\sqrt{n(n+d-1)}\left(\frac{\omega_{n}}{\omega_{n}-1}\right)\psi_{(n-1)mpq}\right\}+\{n\leftrightarrow m\}+\{n\leftrightarrow k\}+\{n\leftrightarrow l\} \nonumber \\
&=-(d-1)\left(\frac{1}{\omega_{n}^{2}-1}+3\right)\psi_{nmpq}+\sqrt{(n+1)(n+d)}\left(\frac{1}{\omega_{n}+1}\right)\psi_{(n+1)mpq} \nonumber \\
&+\sqrt{n(n+d-1)}\left(\frac{1}{\omega_{n}-1}\right)\psi_{(n-1)mpq}
\label{Eqn:HomogeneousEquationPsi}
\end{align}

\subsection{Recursive evaluation of the $X$ and $Y$ coefficients}

One important feature of the recursive expressions in the previous section, is that we can easily compute all the $X$ and $Y$ coeffients recursively. We will explain this in detail for the $X$ coefficients. The procedure is completely analogous for the $Y$ coefficients. 

One starts by evaluating $\chi$ as defined by (\ref{defchi}). For $\chi_{nmpq}$, we call $L=n+m+p+q$ the `level' of that coefficient. At level $0$, there is only one coefficient, which can easily be evaluated analytically,
\begin{equation}
\chi_{0000}=6\frac{\left(\Gamma\left(d\right)\right)^{2}}{\Gamma\left(2d\right)}\frac{\Gamma\left(\frac{3d}{2}\right)}{\left(\Gamma\left(\frac{d}{2}\right)\right)^{3}}.
\end{equation}
Now note that the formula \eqref{Eqn:HomogeneousEquationChi} relates coefficients of level $L-1$ and $L$ to multiple coefficients of level $L+1$. However, by substituting equation \eqref{Eqn:WaveEquationChi} three times in \eqref{Eqn:HomogeneousEquationChi}, we can find a formula that expresses a single coefficient of level $L+1$ in terms of coefficients of level $L$ and $L-1$:
\begin{align}
&\sqrt{(n+1)(n+d)}\frac{(\omega_{n}+\omega_{m}+\omega_{p}+\omega_{q}+2)}{(\omega_{n}+1)}\chi_{(n+1)mpq} \nonumber \\
&=(d-1)\left(\frac{\omega_{m}^{2}}{\omega_{m}-1}+\frac{\omega_{p}^{2}}{\omega_{p}-1}+\frac{\omega_{q}^{2}}{\omega_{q}-1}-(\omega_{m}+\omega_{p}+\omega_{q}+1)\frac{\omega_{n}^{2}}{(\omega_{n}^{2}-1)}\right)\chi_{nmpq} \nonumber \\
&+\sqrt{n(n+d-1)}\frac{(\omega_{n}-\omega_{m}-\omega_{p}-\omega_{q}-2)}{(\omega_{n}-1)}\chi_{(n-1)mpq}+\sqrt{m(m+d-1)}\left(\frac{2\omega_{m}}{\omega_{m}-1}\right)\chi_{n(m-1)pq} \nonumber \\
&+\sqrt{p(p+d-1)}\left(\frac{2\omega_{p}}{\omega_{p}-1}\right)\chi_{nm(p-1)q}+\sqrt{q(q+d-1)}\left(\frac{2\omega_{q}}{\omega_{q}-1}\right)\chi_{nmp(q-1)}.
\end{align}
So once all coefficients $\chi_{nmpq}$ of two adjacent levels are known, one can easily compute all coefficients of the next level. Since the need for explicit integration is completely eliminated, no fast oscillating integrals have to be computed explicitly, and this method will be more efficient than a direct evaluation of the integrals of mode functions. After the recursive computation of the $\chi$ coefficients, one only needs to use \eqref{Eqn:XfromChi} to determine the $X$ coefficients. 

The method for the $Y$ coefficients and the $\psi$ coefficients is completely analogous. One simply needs to combine equations \eqref{Eqn:WaveEquationPsi} and\eqref{Eqn:HomogeneousEquationPsi} and start with the value\footnote{There were typos in this formula in the original version of the paper. We thank Brad Cownden and Andrew Frey for bringing that to our attention.} 
\beq
\psi_{0000}=\frac{8\left(\Gamma\left(d\right)\right)^{2}}{\Gamma\left(2d+1\right)}\frac{\Gamma\left(\frac{3d}{2}-1\right)\Gamma\left(\frac{d}{2}+2\right)}{\left(\Gamma\left(\frac{d}{2}\right)\right)^{4}}.
\eeq

Once all $X$ and $Y$ coefficients are known, we can determine the coefficients $S_{ijkl}$ that appear in the flow equation \eqref{Eqn:RG} for the amplitudes. In order to determine the $R_{ij}$ and $T_{i}$ coefficients, we also need to compute the $W_{ijkl}$ integrals appearing in \eqref{W00} and \eqref{W10}. Though there are much fewer $R$ and $T$ coefficients than $S$ coefficients, and hence optimizing their evaluation is probably less crucial, we shall now explain briefly how they can be efficiently computed as well.

\subsection{Computation of the $W$ coefficients}

Consider the $W$-coefficients as defined in \cite{CEV2},
\begin{equation}
W_{ijkl}^{(a,b)}=\int_{0}^{\pi/2}\text{d}x\,e_{i}^{(a)}(x)e_{j}^{(a)}(x)\mu(x)\nu(x)\int_{0}^{x}\text{d}y\,e_{k}^{(b)}(y)e_{l}^{(b)}(y)\mu(y).
\end{equation}
In appendix D of \cite{CEV2} we established the relations
\begin{subequations}\label{Eqn:WFromXY}
\begin{align}
W_{ijkl}^{(0,1)}-\omega_{k}^{2}W_{ijkl}^{(0,0)}&=X_{kijl},
&W_{ijkl}^{(1,1)}-\omega_{k}^{2}W_{ijkl}^{(1,0)}&=Y_{iljk}, \\
\left(\omega_{k}^{2}-\omega_{l}^{2}\right)W_{ijkl}^{(0,0)}&=X_{lijk}-X_{kijl},
&\left(\omega_{k}^{2}-\omega_{l}^{2}\right)W_{ijkl}^{(1,0)}&=Y_{ikjl}-Y_{iljk}, \\
\left(\omega_{k}^{2}-\omega_{l}^{2}\right)W_{ijkl}^{(0,1)}&=\omega_{k}^{2}X_{lijk}-\omega_{l}^{2}X_{kijl},
&\left(\omega_{k}^{2}-\omega_{l}^{2}\right)W_{ijkl}^{(1,1)}&=\omega_{k}^{2}Y_{ikjl}-\omega_{l}^{2}Y_{iljk}.
\end{align}
\end{subequations}
These allow us to express all $W_{ijkl}^{(a,b)}$ for which $k\neq l$ in terms of $X$ and $Y$ coefficients. However, in the expressions for $R_{ij}$ and $T_{i}$ we need the $W$-coefficients with $k=l$. Assuming that we already know the $\chi$, $\psi$, $X$ and $Y$ coefficients, these can be calculated recursively in the following way.

Consider the integral $\int_{0}^{\pi/2}\text{d}x\,e_{i}e_{j}\mu\nu\int_{0}^{x}\text{d}y\,\mu(\mu\nu')e_{k}e_{l}$. By letting the $(\mu\nu')$ work on $e_{k}$ and $e_{l}$ respectively and using the recurrence relations \eqref{Eqn:Recurrence2e}, we find that
\begin{align}
&(d-1)\left(\frac{1}{\omega_{k}^{2}-1}\right)W_{ijkl}^{(0,0)}+\sqrt{(k+1)(k+d)}\bigg(\frac{W_{ij(k+1)l}^{(0,0)}}{\omega_{k}+1}\bigg)+\sqrt{k(k+d-1)}\bigg(\frac{W_{ij(k-1)l}^{(0,0)}}{\omega_{k}-1}\bigg) \nonumber \\
&=(d-1)\left(\frac{1}{\omega_{l}^{2}-1}\right)W_{ijkl}^{(0,0)}+\sqrt{(l+1)(l+d)}\bigg(\frac{W_{ijk(l+1)}^{(0,0)}}{\omega_{l}+1}\bigg)+\sqrt{l(l+d-1)}\bigg(\frac{W_{ijk(l-1)}^{(0,0)}}{\omega_{l}-1}\bigg)
\label{Eqn:WaveEquationWkl}
\end{align}
Now we substitute $l=k+1$ in this formula and replace all $W_{nmpq}^{(0,0)}$ with $p\neq q$ that appear in the remaining expression by \eqref{Eqn:WFromXY}. What remains is a recursive relation that relates $W_{ijkk}^{(0,0)}$ simply to $W_{i,j,k+1,k+1}^{(0,0)}$. This allows for a fast recursive computation of the $W^{(0,0)}_{ijkk}$ coefficients, once the value of $W^{(0,0)}_{ij00}$ is known.

The most efficient way to compute $W^{(0,0)}_{ij00}$ for all $i$ and $j$ is also by a recursive procedure. In close analogy to the other derivations we have presented, one obtains
\begin{align}
&(d-1)\left(\frac{1}{\omega_{i}^{2}-1}\right)W_{ijkl}^{(0,0)}+\sqrt{(i+1)(i+d)}\bigg(\frac{W_{(i+1)jkl}^{(0,0)}}{\omega_{i}+1}\bigg)+\sqrt{i(i+d-1)}\bigg(\frac{W_{(i-1)jkl}^{(0,0)}}{\omega_{i}-1}\bigg) \nonumber \\
&=(d-1)\left(\frac{1}{\omega_{j}^{2}-1}\right)W_{ijkl}^{(0,0)}+\sqrt{(j+1)(j+d)}\bigg(\frac{W_{i(j+1)kl}^{(0,0)}}{\omega_{j}+1}\bigg)+\sqrt{j(j+d-1)}\bigg(\frac{W_{i(j-1)kl}^{(0,0)}}{\omega_{j}-1}\bigg)
\label{Eqn:WaveEquationWij}
\end{align}
and
\begin{align}
&\left\{\frac{1}{2}(d-1)\left(\frac{\omega_{i}^{2}}{\omega_{i}^{2}-1}\right)W_{ijkl}^{(0,0)}-\frac{1}{2}\sqrt{(i+1)(i+d)}\left(\frac{\omega_{i}}{\omega_{i}+1}\right)W_{(i+1)jkl}^{(0,0)}\right. \nonumber \\
&\left.+\frac{1}{2}\sqrt{i(i+d-1)}\left(\frac{\omega_{i}}{\omega_{i}-1}\right)W_{(i-1)jkl}^{(0,0)}\right\}+\{i\leftrightarrow j\}+\{i\leftrightarrow k\}+\{i\leftrightarrow l\} \nonumber \\
&=(d-1)W_{ijkl}^{(0,0)}+2(d-1)\left(\frac{1}{\omega_{i}^{2}-1}\right)W_{ijkl}^{(0,0)}+2\sqrt{(i+1)(i+d)}\left(\frac{1}{\omega_{i}+1}\right)W_{(i+1)jkl}^{(0,0)} \nonumber \\
&+2\sqrt{i(i+d-1)}\left(\frac{1}{\omega_{i}-1}\right)W_{(i-1)jkl}^{(0,0)}+(d-1)\left(\frac{1}{\omega_{k}^{2}-1}\right)W_{ijkl}^{(0,0)} \nonumber \\
&+\sqrt{(k+1)(k+d)}\left(\frac{1}{\omega_{k}+1}\right)W_{ij(k+1)l}^{(0,0)}+\sqrt{k(k+d-1)}\left(\frac{1}{\omega_{k}-1}\right)W_{ij(k-1)l}^{(0,0)}
\label{Eqn:HomogeneousEquationW}
\end{align}
Using these relations, the $W^{(0,0)}_{ij00}$ coefficients can be computed in exactly the same way as for the $\chi$ and $\psi$ coefficients. Only now we take $L=i+j$ to be the `level' of $W^{(0,0)}_{ij00}$.

We can easily extract the $W^{(1,0)}$ coefficients once all the $W^{(0,0)}$ are known. On the one hand, we consider $W_{ijkl}^{(0,0)}$, with $e_i$ expressed as $-(\mu e'_{i})'/\om_i^2$, and use integration by parts to throw one of the two thus inserted derivatives off $e_i$, obtaining the relation
\beq
-\omega^{2}_{i}W_{ijkl}^{(0,0)}
=-W_{ijkl}^{(1,0)}-N_{ijkl}^{(0)}-X_{ijkl}
\eeq
where we have defined the integral $N_{ijkl}^{(a)}=\int_{0}^{\pi/2}\text{d}x\,e'_{i}e_{j}\mu\nu'\int_{0}^{x}\text{d}y\,e_{k}^{(a)}e_{l}^{(a)}\mu$. On the other hand, we have by integration by parts, 
\beq
N_{ijkl}^{(0)}+N_{jikl}^{(0)}
=4W_{ijkl}^{(0,0)}+2(d-1)\chi_{ijkl}+X_{ijkl}+X_{jikl}+X_{kijl}+X_{lijk}.
\eeq
Here we have used the fact that $(\mu\nu')'=-4\mu\nu$ and the expression for the integral \eqref{Eqn:IntegralChiN}. Combining these two, we find that
\begin{equation}
(\omega^{2}_{i}+\omega^{2}_{j}-4)W_{ijkl}^{(0,0)}-2W_{ijkl}^{(1,0)}=2(d-1)\chi_{ijkl}+2X_{ijkl}+2X_{jikl}+X_{kijl}+X_{lijk}
\end{equation}
Assuming all $\chi$, $X$ and $W^{(0,0)}$ are known, this allows us to compute all $W^{(1,0)}$.

\subsection{Additional coefficients in interior time gauge}

If one works in the interior time ($\delta(0,t)=0$), one also needs the $V_{ij}$ and $A_{ij}$ coefficients defined by (\ref{VAdef}). These coefficients can also be computed by a simple recursive procedure. 

Just as before, we first consider the action of the recursion relation (\ref{Eqn:Recurrence1e}) either on $\mu\nu'e_n$ or on $\mu\nu'e_m$ in the product  $\mu\nu'e_ne_m$. We then equate the results, multiply by $\mu\nu$ and integrate to obtain
\begin{align}
&\frac{(d-1)}{\omega_{n}^{2}-1}V_{nm}+\sqrt{(n+1)(n+d)}\left(\frac{V_{(n+1)m}}{\omega_{n}+1}\right)+\sqrt{n(n+d-1)}\left(\frac{V_{(n-1)m}}{\omega_{n}-1}\right) \nonumber \\
&=\frac{(d-1)}{\omega_{m}^{2}-1}V_{nm}+\sqrt{(m+1)(m+d)}\left(\frac{V_{n(m+1)}}{\omega_{m}+1}\right)+\sqrt{m(m+d-1)}\left(\frac{V_{n(m-1)}}{\omega_{m}-1}\right).
\end{align}
We then note that
\begin{align}
&(d-1)V_{ij}=(d-1)\int_{0}^{\frac{\pi}{2}}\text{d}x\,e_{i}e_{j}\mu\nu=\int_{0}^{\frac{\pi}{2}}\text{d}x\,e_{i}e_{j}\mu\nu(\mu'\nu)=-\int_{0}^{\frac{\pi}{2}}\text{d}x\,(e_{i}e_{j}\mu\nu^{2})'\mu \nonumber \\
&=-\int_{0}^{\frac{\pi}{2}}\text{d}x\,(\mu\nu e'_{i})e_{j}\mu\nu-\int_{0}^{\frac{\pi}{2}}\text{d}x\,e_{i}(\mu\nu e'_{j})\mu\nu-\int_{0}^{\frac{\pi}{2}}\text{d}x\,e_{i}e_{j}\mu\nu(\mu'\nu)-2\int_{0}^{\frac{\pi}{2}}\text{d}x\,(\mu\nu'e_{i})e_{j}\mu\nu
\end{align}
Using the recurrence relations results in
\begin{align}
&\left\{\frac{1}{2}(d-1)\left(\frac{\omega_{n}^{2}}{\omega_{n}^{2}-1}\right)V_{nm}-\frac{1}{2}\sqrt{(n+1)(n+d)}\left(\frac{\omega_{n}}{\omega_{n}+1}\right)V_{(n+1)m}\right. \nonumber \\
&\left.+\frac{1}{2}\sqrt{n(n+d-1)}\left(\frac{\omega_{n}}{\omega_{n}-1}\right)V_{(n-1)m}\right\}+\{n\leftrightarrow m\} \nonumber \\
&=2(d-1)\left(\frac{1}{\omega_{n}^{2}-1}\right)V_{nm}+2\sqrt{(n+1)(n+d)}\left(\frac{1}{\omega_{n}+1}\right)V_{(n+1)m} \nonumber \\
&+2\sqrt{n(n+d-1)}\left(\frac{1}{\omega_{n}-1}\right)V_{(n-1)m}
\end{align}

The above relations can be used to efficiently compute the $V_{ij}$ coefficients, starting with 
\beq
V_{00}=\frac{2\,\Gamma(d)}{(d+1)\left(\Gamma\left(\frac{d}{2}\right)\right)^{2}}.
\eeq 
In appendix B of \cite{CEV2}, we established that
\begin{equation}
A_{ij}=\frac{1}{2}(\omega_{i}^{2}+\omega_{j}^{2}-4)V_{ij}-\frac{1}{2}C_{i}C_{j}
\qquad\text{with}\qquad
C_{i}\equiv\frac{2\sqrt{d-2}}{\Gamma(d/2)}\sqrt{\frac{(i+d-1)!}{i!}}.
\end{equation}
Therefore, we can immediately obtain all $A$-coefficients as well.

\section{Ultraviolet asymptotics of the interaction coefficients}

We can extract the behavior of the interaction coefficients for large mode numbers by solving the recurrence relations asymptotically in this limit. In particular, for the $\chi_{ijkl}$ coefficients, we consider the recurrence relations \eqref{Eqn:WaveEquationChi} and \eqref{Eqn:HomogeneousEquationChi}. In the limit of large values of $n$, $m$, $p$ and $q$, these reduce to
\beq
(d-1)\chi_{nmpq}+\frac{1}{2}\chi_{n(m+1)pq}+\frac{1}{2}\chi_{n(m-1)pq}=(d-1)\chi_{nmpq}+\frac{1}{2}\chi_{(n+1)mpq}+\frac{1}{2}\chi_{(n-1)mpq}
\eeq
and
\begin{align}
&\left\{\frac{1}{2}(d-1)\chi_{nmpq}-\frac{1}{2}n\chi_{(n+1)mpq}+\frac{1}{2}n\chi_{(n-1)mpq}\right\}+(n\leftrightarrow m)+(n\leftrightarrow p)+(n\leftrightarrow q) \nonumber \\
&=(d-1)\chi_{nmpq}+\frac{1}{2}\chi_{(n+1)mpq}+\frac{1}{2}\chi_{(n-1)mpq}.
\end{align}

By introducing finite difference operators $\Delta^{\hspace{-.8mm}(1)}_{n}\chi_{nmpq}=(\chi_{(n+1)mpq}-\chi_{(n-1)mpq})/2$ and $\Delta^{\hspace{-.8mm}(2)}_{n}\chi_{nmpq}=\chi_{(n+1)mpq}+\chi_{(n-1)mpq}-2\chi_{nmpq}$ , we can write these equations as
\begin{equation}
\Delta^{\hspace{-.8mm}(2)}_{m}\chi_{nmpq}=\Delta^{\hspace{-.8mm}(2)}_{n}\chi_{nmpq}
\label{discrwave}
\end{equation}
and
\begin{equation}
n\Delta^{\hspace{-.8mm}(1)}_{n}\chi_{nmpq}+m\Delta^{\hspace{-.8mm}(1)}_{m}\chi_{nmpq}+p\Delta^{\hspace{-.8mm}(1)}_{p}\chi_{nmpq}+q\Delta^{\hspace{-.8mm}(1)}_{q}\chi_{nmpq}=(d-1)\chi_{nmpq}-\frac12(\chi_{(n+1)mpq}+\chi_{(n-1)mpq}).
\label{discrhom}
\end{equation}
Note that our finite difference operators are simply proportional to discretized first and second derivatives.
(\ref{discrwave}) is a discretized two-dimensional wave equation and can be solved by an ansatz of the following form (which takes into account the total symmetry of $\chi_{nmpq}$):
\begin{align}
\chi_{nmpq}=&f_{1}(n+m+p+q)+f_{2}(-n+m+p+q)+f_{2}(n-m+p+q) \nonumber \\
&+f_{2}(n+m-p+q)+f_{2}(n+m+p-q) \nonumber \\
&+f_{3}(n+m-p-q)+f_{3}(n-m+p-q)+f_{3}(n-m-p+q),
\end{align}
where the function $f_3$  must be even, $f_{3}(-x)=f_{3}(x)$. One can develop some intuition about (\ref{discrhom}) by likewise considering its continuum limit, assuming $|\chi_{n+1}-\chi_{n}|\ll |\chi_n|$, which yields the Euler equation for homogeneous functions,
\begin{equation}
n\partial_{n}\chi+m\partial_{m}\chi+p\partial_{p}\chi+q\partial_{q}\chi=(d-2)\chi,
\end{equation}
solved by taking $f_{i}(x)=A_{i}x^{d-2}$. (For $d=2$, we expect a logarithmic behavior $f_{i}(x)=A_{i}\ln x$.) 
One can directly verify that such a solution is consistent with the original finite difference equation (\ref{discrhom}) at large values of the indices. Overall,
\begin{align}
&\chi_{nmpq}\sim A_{3}\left((n+m-p-q)^{d-2}+(n-m+p-q)^{d-2}+(n-m-p+q)^{d-2}\right)\nonumber \\
&\hspace{1cm}+A_{2}\left((-n+m+p+q)^{d-2}+(n-m+p+q)^{d-2}+(n+m-p+q)^{d-2}\right.\nonumber \\
&\hspace{2cm}+\left.(n+m+p-q)^{d-2}\right)+A_{1}(n+m+p+q)^{d-2}.
\end{align}
(We have compared this expression to some explicit brute force calculations for special index values using Mathematica.)

One can check that the asymptotic equations for the $\psi_{nmpq}$ coefficients are exactly the same, thus we expect the same kind of behavior. Extending this to the $X$ and $Y$ coefficients, plugging this back in \eqref{Eqn:Sijkl} and assuming that there are no non-trivial cancellations, we expect the following scaling
\begin{equation}
S_{\lambda i,\lambda j,\lambda k,\lambda l}\sim\lambda^{d},
\label{Sscale}
\end{equation}
for any $i,j,k,l$, with $\lambda\gg 1$.

Similar considerations yield for the $W$-coefficients
\beq
W^{(0,0)}_{\lambda i,\lambda j,\lambda k,\lambda l}\sim\lambda^{d-4},\qquad W^{(1,0)}_{\lambda i,\lambda j,\lambda k,\lambda l}\sim\lambda^{d-2}.
\eeq
Hence, the $R$ and $T$-coefficients in the {\em boundary} time gauge behave as
\beq
R_{\lambda i,\lambda j}\sim\lambda^{d},\qquad T_{\lambda i}\sim\lambda^{d}.
\label{RTassymp}
\eeq

Note that the $R$ and $T$ coefficients in the {\em interior} time gauge feature extra terms, as explained above (\ref{VAdef}). Repeating the asymptotic analysis for the $V_{ij}$ coefficients, we find the equations
\begin{align}
n\Delta^{\hspace{-.8mm}(1)}_{n}V_{nm}+ m\Delta^{\hspace{-.8mm}(1)}_{m}V_{nm}=(d-3)V_{nm}
\end{align}
and
\begin{equation}
\Delta^{\hspace{-.8mm}(2)}_{n}V_{nm}=\Delta^{\hspace{-.8mm}(2)}_{m}V_{nm}
\end{equation}
which are solved by $V_{nm}\sim\nu_{1}(n+m)^{d-3}+\nu_{2}(n-m)^{d-3}$. For $d=3$, this goes to $\ln(n+m)$ and $\ln(n-m)$ instead, and for $d=2$, this goes to a constant (the explicit value of this constant is given in appendix B of \cite{CEV2}).
The dominant contribution to the $T$ and $R$-coefficients becomes $\om_i^4 V_{ii}$ and $\om_i^2\om_j^2V_{ii}$, respectively, rather than coming from the terms already accounted for in (\ref{RTassymp}), which become subleading in the interior gauge. For AdS$_4$ this results in the behavior $T_j\sim j^4\ln j$ and $R_{jn}\sim n^2j^2\ln j$, consistent with the one reported in \cite{GMLL}. For AdS$_{d+1}$ with $d>3$, it results in the behavior  $T_j\sim j^{d+1}$ and $R_{jn}\sim n^2j^{d-1}$, consistent with the one reported in \cite{BMR} from numerical evaluations in AdS$_5$. Note that we do not find evidence for logarithmic modulations in the leading power-law behavior of the interior gauge interaction coefficients for $d>3$ (a possibility suggested in footnote 9 of \cite{GMLL}).

\section{Discussion}

We have presented some analytic considerations of the interaction coefficients appearing in the time-averaged equations describing slow energy transfer between the normal modes of a spherically symmetric AdS-scalar field system due to gravitational non-linearities. In the course of our considerations, we have developed an iterative procedure for evaluating the interaction coefficients, which is likely to present significant advantages over a brute force approach. Indeed, straightforward computation of the interaction coefficients can be done in essentially two ways, both of which are resource-intensive. Either one multiplies high-degree Jacobi polynomials symbolically and integrates the result exactly using computer algebra, or one resorts to a numerical treatment of the integrals of products of Jacobi polynomials, which become very rapidly oscillating as the mode number increases. Our recursive procedure eliminates the need for integration altogether (the only drawback is that one has to evaluate the interaction coefficients for all modes, rather than only for those that satisfy the resonance condition). This sort of optimization is likely to enable significant improvements in the numerical analysis of the time-averaged theory of the sort presented in \cite{BMR}. It would also be interesting to know how the iterative evaluation competes in terms of computational efficiency with the explicit expressions for the interaction coefficients in the particular case of AdS$_4$ presented in \cite{GMLL}, as those expressions are extremely long and require many floating point operations per coefficient.

Our iterative formulas allow us to deduce the asymptotic power law dependences of the interaction coefficients. In the boundary time gauge, if all the mode numbers are scaled up uniformly controlled by a parameter $\lambda$, all interaction coefficients scale as $\lambda^d$ in AdS$_{d+1}$. The ultraviolet growth of mode couplings becomes more harsh in higher dimensions, conforming to the strengthening of turbulent behavior in higher dimensions alluded to in \cite{GMLL}.

It would be interesting to explore how the power law (\ref{Sscale}) appearing in the scaling of the $S$ coefficients relates to the properties of the singular spectra developing in finite time in solutions of the time-averaged system according to \cite{BMR}. (This singular behavior is believed to reflect the black hole formation in the full non-linear theory, though establishing this relation rigorously would require going beyond the time-averaged description.) In \cite{BMR}, it was observed numerically that, in AdS$_5$, the following behavior of the amplitude spectrum $A_n$ emerges near a certain finite `singular' moment $t_*$:
\beq
A_n(\eps^2 t)\sim n^{-2} e^{-\rho\hspace{.7pt} n\hspace{.7pt} \eps^2(t_*-t)}.
\eeq
This solution was verified to be crudely analytically consistent with the time-averaged amplitude equation (\ref{Eqn:RG}) in the sense that both sides appear approximately independent of $n$ at large $n$, given the scaling $S_{\lambda i,\lambda j,\lambda k,\lambda l}\sim\lambda^{4}$ in AdS$_5$. We hope that our results, when combined with generalizations of the numerical analysis of  \cite{BMR} to higher dimensions, will lead to a more detailed and more general understanding of such singular spectra.

Another comment is that in \cite{BMR}, which used interior time gauge, the flow equations for $d=4$  led to finite-time singular behavior of (the time-derivative of) the phases $B_n$, and this singular behavior seemed to be driven by the $R_{jn}\sim n^2j^{3}$ scaling. Our analysis shows that this scaling is due to a contribution that is absent in boundary time gauge, so it would be interesting to verify whether the singular behavior of the phases persists in boundary time gauge.


\section{Acknowledgments}

We would like to thank Piotr Bizo\'n, Chethan Krishnan, Luis Lehner, Maciej Maliborski and Andrzej Rostworowski for useful discussions. The work of B.C.\ and J.V.\ has been supported by the Belgian Federal Science Policy Office through the Interuniversity Attraction Pole P7/37, by FWO-Vlaanderen through project G020714N, and by the Vrije Universiteit Brussel through the Strategic Research Program ``High-Energy Physics.'' The work of O.E. is funded under CUniverse research promotion project by Chulalongkorn University (grant reference CUAASC). J.V.\ is supported by a PhD fellowship of Research Foundation Flanders (FWO).


\end{document}